\documentclass[10pt,conference]{IEEEtran}
 \usepackage{amssymb}
 \usepackage{amsmath}
\usepackage{graphicx} 
\usepackage{lipsum}
\usepackage{tikz}
\usetikzlibrary{arrows.meta}
\usetikzlibrary{arrows,shapes,snakes,automata,backgrounds,petri,positioning,decorations.markings}
\usepackage{quantikz}
\usepackage{cases}
\usepackage{subcaption}
\usepackage{cite}
\usepackage{hyperref}
\usepackage{multirow, booktabs}
\usepackage{algorithm, algpseudocode}

\newcommand{\intset}[1]{[\![ #1 ]\!]}
\newcommand{\qed}{\hfill $\diamond$}

\newtheorem{example}{Example}
\newtheorem{remark}{Remark}
\newtheorem{proposition}{Proposition}

\title{Optimizing Variational Circuits for\\ Higher-Order Binary Optimization}

\author{
\IEEEauthorblockN{Zoé Verchère$^{1, 2}$, Sourour Elloumi$^{1, 2}$, Andrea Simonetto$^1$}
\vspace*{2mm}
\IEEEauthorblockA{$^1$ Unité de Mathématiques Appliquées, ENSTA Paris, Institut Polytechnique de Paris, 91120 Palaiseau, France}
\IEEEauthorblockA{$^2$ CEDRIC,  Conservatoire National des Arts et M\'etiers, 75003 Paris, France}
\vspace*{-5mm}}

\date{March 2023}

\begin{document}

\maketitle

\begin{abstract}
   Variational quantum algorithms have been advocated as promising candidates to solve combinatorial optimization problems on near-term quantum computers. Their methodology involves transforming the optimization problem into a quadratic unconstrained binary optimization (QUBO) problem. While this transformation offers flexibility and a ready-to-implement circuit involving only two-qubit gates, it has been shown to be less than optimal in the number of employed qubits and circuit depth, especially for polynomial optimization. On the other hand, strategies based on higher-order binary optimization (HOBO) could save qubits, but they would introduce additional circuit layers, given the presence of higher-than-two-qubit gates. 

   In this paper, we study HOBO problems and propose new approaches to encode their Hamiltonian into a ready-to-implement circuit involving only two-qubit gates. Our methodology relies on formulating the circuit design as a combinatorial optimization problem, in which we seek to minimize circuit depth. We also propose handy simplifications and heuristics that can solve the circuit design problem in polynomial time. We evaluate our approaches by comparing them with the state of the art, showcasing clear gains in 
   terms of circuit depth.     
\end{abstract}


\section{Introduction}

Variational quantum algorithms are being investigated as a promising strategy to solve well-known combinatorial optimization problems on near-term quantum devices. An example of such algorithms is QAOA \cite{QAOA,peruzzo2014variational}. The original paper by Farhi, Goldstone and Gutmann applies this method to Max-Cut, a notoriously NP-hard combinatorial problem. From there, it was quickly understood that QAOA can be used to approximately solve any Quadratic Unconstrained Binary Optimization (QUBO) problems. This understanding established a simple workflow: start from a given optimization problem, reformulate it as a QUBO, then use QAOA. In \cite{L14}, Lucas gives QUBO reformulations of many well-known NP-complete and NP-hard problems, including the satisfiability (SAT) problem, the graph isomorphism problem, and the traveling salesman problem (TSP). Since then, many variants and applications have appeared~\cite{mcclean2016theory, farhi2017quantum,Zhou2018, barkoutsos2019improving,nannicini2019performance,gambella2020multi,chatterjee2023solving} (and references therein). Whenever higher-order terms appear, they are reduced to quadratic terms by the introduction of auxiliary variables. In \cite{GOQAOA}, Herrman et al. discuss how to introduce such auxiliary variables in a way that guarantees that the circuit encoding the resulting QUBO is as shallow as possible.

Even though the QUBO form is quite practical and versatile, recent work by Glos, Krawiec and Zimbor{\`a}s \cite{HOBO} shows that reformulating the original problem at hand as a higher-order binary optimization (HOBO) problem can be interesting in a number of ways. Working on the TSP, they show that considering HOBO reformulations allows for different compromises between number of qubits and circuit depth needed to encode the Hamiltonian corresponding to the problem. In the case of the TSP, they manage to encode a HOBO reformulation on fewer qubits than the state-of-the-art QUBO reformulation ($\mathcal{O}(N \log(N))$ qubits for $N$ binary variables, against $\mathcal{O}(N^2)$), but at the cost of a deeper circuit.

This increased depth is in part due to the current available technology. In the QUBO case, there is a rather clear workflow between the QUBO formulation of the problem and the parametrized circuit for the corresponding Hamiltonian. A change of variable lets us change from binary variables into convenient Pauli operators. The corresponding formulation of the problem is often called the Ising formulation, due to the similarities with a model called Ising spin glass in existing literature \cite{L14,IsingComp}. This formulation then informs us on how to build the circuit moving forward, using only two-qubit gates. In the case of a HOBO formulation, the corresponding Ising formulation is also of degree greater than two, implying the use of $d$-qubit gates. Such gates are hard to implement, and must therefore be decomposed into a series of smaller gates, therefore making the circuit deeper.

In this paper, we propose new ways to design variational circuits for HOBO problems. We operate under strict hypotheses as to which gates may be used, thus ensuring the resulting circuit may directly be implemented on currently available technology. Our contributions are as follows,

$\bullet$ We reinterpret and formulate the circuit design as a combinatorial routing problem, that we model as a mixed-integer linear programming problem, which can minimize the circuit depth;

$\bullet$ We propose simplifications and heuristics to approximate the solution of the combinatorial design problem, so to derive the variational circuit in polynomial time;

$\bullet$ The proposed heuristic is a recursive circuit template, computable in polynomial time, that can be used to synthetize the circuit of any HOBO problem of $m$ monomials and degree $D$, in $O(m\, 2^D)$ layers;

$\bullet$ We demonstrate the performance of the proposed design methodologies, comparing them with the state of the art, highlighting clear gains in terms of circuit depth.  

\smallskip

{\bf Organization.} Sect.~\ref{sec:formulation} presents the problem formulation. We derive the circuit design problem in Sect.~\ref{sec:design}, and we propose simplifications and heuristics in Sect.~\ref{sec:solutions}. Sec.~\ref{sec:results} collects our numerical comparisons, and we close in Sect.~\ref{sec:conclusion}.

\section{Problem formulation}\label{sec:formulation}

Our ultimate goal is to solve the following HOBO problem:
\begin{equation}
    \underset{x \in \{0,1\}^n}{\min} f(x) := \underset{M \in P}{\sum} \Big[C_M \underset{i \in M}{\prod} x_i\Big],
\end{equation}
where $P$ is the polynomial set of all monomials $M$, $C_M \in \mathbb{R}$ for all $M$ are the monomial coefficients, and $x_i$'s are the decision variables. In this context, a monomial $M$ is the set of indices of the variables appearing in the corresponding term of the objective function. Since the variables are binary, $x_i = x_i^2$ for all $i$, therefore we need not consider adding exponents on any of the variables. The degree of a monomial $M$ is thus equal to its cardinality, $|M|$. The overall degree of the function $f$ is ${\max_{M \in P}} |M|$. If the degree of $f$ is equal to two, then this problem falls into the QUBO category. If it is greater than two, then it falls into the HOBO category, which we plan to study.

The first step towards using a quantum variational algorithm to solve this problem is to derive its (Ising) Hamiltonian. This can be done by using the variable change $x_i \leftarrow \frac{1-Z_i}{2}$ for all $i$, where $Z_i$ is the Pauli operator that acts on qubit $i$. This gives the following diagonal Hamiltonian,
\begin{equation}\label{eq:hamiltonian}
     H(Z) := \underset{M \in P_H}{\sum} \Big[\alpha_M \underset{i \in M}{\prod} Z_i\Big],
\end{equation}
where $P_H$ represents the polynomial set in the new variables and $\alpha_M$ are the new real-valued coefficients. 

We notice that $P_H$ is constituted of all the monomials which are subsets of at least one of the monomials of $P$, the polynomial in binary variables. This is due to the variable change: let us briefly focus on the case of a single monomial of degree three, $f(x) = x_1 x_2 x_3$. Then, the corresponding Ising Hamiltonian is $H(Z) = (\frac{1-Z_1}{2})(\frac{1-Z_2}{2})(\frac{1-Z_3}{2}) = \frac{1}{8}(1-Z_1-Z_2-Z_3+Z_1Z_2+Z_1Z_3+Z_2Z_3-Z_1Z_2Z_3)$, creating every subset of the original monomial into monomials in the Ising Hamiltonian.

Given the Hamiltonian $H(Z)$ in~\eqref{eq:hamiltonian}, the goal of this paper is to study how we can encode it into a variational quantum circuit while minimizing circuit depth.

\section{Circuit design as routing}\label{sec:design}

We start by noticing that in variational algorithms, such as QAOA, we do not encode the Ising Hamiltonian $H(Z)$ directly, but we actually encode an operator: a $2^n \times 2^n$ unitary diagonal matrix defined as 
\begin{equation}
  U_H(\gamma) =  \mathrm{e}^{-\jmath \gamma H(Z)} = \exp\Big(-\jmath \gamma \underset{M \in P_H}{\sum} \alpha_M \underset{i \in M}{\prod} {Z}_i\Big),  
\end{equation}
where $\gamma$ is a free rotation parameter that will be optimized over classically, and $\jmath = \sqrt{-1}$. In order to design circuits which may be run on currently available (superconducting) technology, we place ourselves under strict conditions concerning available gates. We may use only one-qubit Pauli and phase rotation gates, as well as two-qubit CNOT gates, \cite{madden2022best}. Without loss of generality, for circuit design purposes only, we let $\gamma=1$. 
\smallskip

\begin{figure*}
\centering
    \begin{subfigure}{0.415\textwidth}
    \centering
    \includegraphics[width=0.8\textwidth]{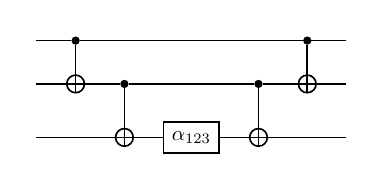}
    \subcaption{Circuit for $U_H = \exp(-\jmath\hspace{1mm}\alpha_{123}Z_1 Z_2 Z_3)$ without ancilla.}
    \end{subfigure}
    \hfill 
    \begin{subfigure}{0.415\textwidth}
    \includegraphics[width=1\textwidth]{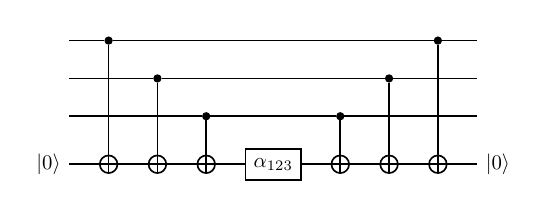}
    \subcaption{Circuit for $U_H = \exp(-\jmath \hspace{1mm} \alpha_{123}Z_1 Z_2 Z_3)$ with an ancilla.}
    \end{subfigure}
    \begin{subfigure}{0.78\textwidth}
    \centering
    \includegraphics[width=\textwidth]{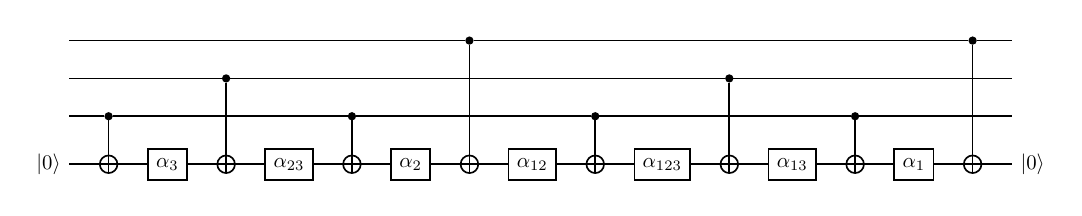}
    \subcaption{Circuit for $U_H = \exp(-\jmath \underset{I \subset \{1,2,3\}}{\sum}\alpha_I \underset{i \in I}{\prod}Z_i)$, using Gray code and an ancilla.}
    \end{subfigure}
    \caption{Quantum circuits for a few simple Ising Hamiltonians.}
    \label{fig:decomp_example}
\end{figure*}

\begin{example}
To fix the ideas, let us consider a very simple Ising Hamiltonian, $H(Z) = \alpha_{123} Z_1 Z_2 Z_3$. The corresponding operator is $U_H = \exp(-\jmath\alpha_{123}Z_1 Z_2 Z_3)$. A known \cite{HOBO,BK12} decomposition of this operator using the allowed gates is depicted in Figure \ref{fig:decomp_example}a. The rotation can also be applied on a provided ancilla qubit as in Figure \ref{fig:decomp_example}b. \qed
\end{example}

\smallskip
\begin{example}
Next, we consider a slightly more complicated Ising Hamiltonian of the following form: 
$$
H(Z) = \underset{I \subset \{1,2,3\}}{\sum} [ \alpha_I  \underset{i \in I}{\prod} Z_i].
$$ 
This Hamiltonian stems from the variable change applied to $f(x) = x_1 x_2 x_3$. Inspired by Gray code (also known as reflected binary code), in \cite{HOBO}, the authors propose the circuit shown in Figure \ref{fig:decomp_example}c. Since this last circuit lets us encode a monomial of degree three of the original binary formulation of the problem, a first possible design for the whole circuit is to use this method to generate a circuit for each monomial, and then put them together end-to-end to obtain a full circuit. However, this has two obvious downsides.

The first downside is that we may end up repeating some elements over the course of the whole circuit. For example, let us consider $f(x) = C_{123} x_1 x_2 x_3 + C_{234} x_2 x_3 x_4$. Using this method based on Gray code to build one subcircuit per monomial means that we are building a circuit that encodes $H_1(Z) = \underset{I \subset \{1,2,3\}}{\sum} (-1)^{|I|} C_{123} \prod_{i \in I} Z_i$, and a second circuit that encodes $H_2(Z) = \underset{I \subset \{2,3,4\}}{\sum} (-1)^{|I|} C_{234} \prod_{i \in I} Z_i$, and putting them together one after the other. Therefore, we will include twice all the monomials that are common to the two Ising Hamiltonians, in this case $\{2\}$, $\{3\}$, and $\{2,3\}$.

The second downside is that every phase rotation happens on the ancilla qubit, which means that no two phase rotations can happen at the same time. We believe it is a misuse of resources, and the same operator could be built on a much shallower circuit by parallelizing operations. \qed
\end{example}

\smallskip
From the discussion of the above examples, it seems reasonable that a better strategy to encode the Hamiltonian should be achievable. In this paper, we take the point of view of \emph{routing}. In fact, from a more high-level abstraction, the circuit in Figure \ref{fig:decomp_example}a can be seen as a form of routing of the singletons $Z_i$, in order to form the product $Z_1 Z_2 Z_3$ on the third qubit, at which moment, the appropriate phase rotation gate is applied. 

Following this point of view, at the start of the circuit, each qubit represents its corresponding singleton: $Z_1$ is on qubit 1, $Z_2$ on qubit 2, $Z_3$ on qubit 3, and so forth. The first CNOT gate from qubit 1 to qubit 2 replaces the current monomial on qubit 2 by the product of the monomials on qubit 1 and 2, that is to say $Z_2$ is replaced by $Z_1 Z_2$.

This interpretation is also valid for the circuits in Figures~\ref{fig:decomp_example}b and~\ref{fig:decomp_example}c, where we simply have to consider that the ancilla qubit starts by representing the constant $1$. 

This view led us to conceptualize the search for the shallowest circuit for a given Ising Hamiltonian as a combinatorial optimization problem, where variables and constraints describe which gates are put on which qubits, and this routing process of the monomials through the circuit through the use of the CNOT gates. Let us introduce the following variables and notations that will help us in modeling our routing problem.

\smallskip
\begin{remark}
As the reader may be aware of, QAOA involves also a mixing Hamiltonian, and an expected value computation. Both can be done as usual. The mixing Hamiltonian can be designed to contain two-qubit gates only, and since the problem Hamiltonian is diagonal, then the expected value computation can be performed as usual. Hence, the complicating factor for HOBO problems is the encoding of $U_H$. 
\end{remark}

\subsection{Preliminaries and notations}

As previously mentioned, $n$ is the number of variables in the problem (and its corresponding Ising formulation), and $P_H$ is the set of monomials in the Ising formulation of the problem. The total number of available qubits is denoted by $q$. It verifies $q \geqslant n$, and any qubits beyond the $n$ first are ancillas. The maximum circuit depth considered is called $T$. This parameter may be set to the depth of a known circuit constructed via a simple heuristic. For handy notation we set $\intset{l,m} :=\{l, \ldots, m\}$ for any integer $l\leq m$. 


We introduce the following supporting binary variables:
\begin{itemize}
    \item $a^k$ : 1 if something happens in the circuit on layer $k$, 0 otherwise.
    \item $c^k_{i,j}$ : 1 if there is a CNOT gate from qubit $i$ to $j$ on layer $k$, 0 otherwise.
    \item $r^k_{i,v}$ : 1 if there is a rotation gate corresponding to monomial $v$ on qubit $i$ on layer $k$, 0 otherwise.
    \item $d^k_{i,v}$ : 1 if monomial $v$ is represented on qubit $i$ on layer $k$, after applying the gates for that step, 0 otherwise.
    \item $b^k_{i,p}$ : 1 if $Z_p$ is on qubit $i$ on layer $k$, after applying the gates for that step, 0 otherwise.
\end{itemize}


Since the goal is to reduce circuit depth as much as possible, we must minimize the sum of active layers, so, we introduce the following cost function
\begin{equation}
J(\{a^k\}) = \underset{k=1}{\overset{T}{\sum}} a^k.
\end{equation}

We look now at the routing constraints. 

\textbf{Initial conditions of the circuit:} the first $n$ qubits represent $Z_i$, for all $i \in \{1,\dots,n\}$. Therefore, the following constraints are necessary at the start of the circuit, i.e. layer zero,
\begin{equation}
    \begin{cases}
    b^0_{i,i} = 1 & \forall i \in \intset{1,n}\\
    b^0_{i,p} = 0 & \forall i \in \intset{1,n}, \hspace{2mm} \forall p \neq i\\
    b^0_{i,p} = 0 & \forall i \in \intset{n+1,q}, \hspace{2mm} \forall p \in \intset{1,n}.
    \end{cases}
    \label{eq:init}
\end{equation}

\textbf{Final conditions of the circuit:} uncomputing is required, as this circuit will be part of a bigger circuit. Therefore, the following constraints are necessary at time step $T$,
\begin{equation}
    \begin{cases}
    b^T_{i,i} = 1 & \forall i \in \intset{1,n}\\
    b^T_{i,p} = 0 & \forall i \in \intset{1,n}, \hspace{2mm} \forall p \neq i\\
    b^T_{i,p} = 0 & \forall i \in \intset{n+1,q}, \hspace{2mm} \forall p \in \intset{1,n}.
    \end{cases}
\end{equation}

\textbf{CNOT or rotation implies activity, and uniqueness:} a layer is active if there is a CNOT gate or rotation gate on that layer, and a qubit cannot be acted on by more than one gate per layer. Therefore, we have the following constraints.
\begin{multline}
    \underset{j=1}{\overset{q}{\sum}} (c^k_{i,j} + c^k_{j,i}) + \underset{v \in P_H}{\sum} r^k_{i,v} \leqslant a^k, \, \forall k \in \intset{1,T},\hspace{1mm} \forall i \in \intset{1,q}.
\end{multline}

\textbf{No self-control:} a qubit may not control itself through the use of a CNOT gate,
\begin{equation}
    c^k_{i,i} = 0, \quad \forall k \in \intset{1,T}, \forall i \in \intset{1,q}.
\end{equation}

\textbf{Compute all monomials:} all the monomials in the formulation must be treated by the appropriate rotation exactly once,
\begin{equation}
    \underset{k=1}{\overset{T}{\sum}} \underset{i=1}{\overset{q}{\sum}} r^k_{i,v} = 1 \quad \forall v \in P_H .
\end{equation}

\textbf{Monomial check:} check whether a given monomial of $P_H$ is represented on a given qubit at a given time step,
\begin{equation}
    \begin{cases}
    d^k_{i,v} \leqslant b^k_{i,p} & \\ &\hspace*{-4cm}\forall k \in \intset{0,T}, \forall i \in \intset{1,q}, \forall v \in P_H, \forall p \in v \\
    d^k_{i,v} \leqslant 1 - b^k_{i,p} & \\ &\hspace*{-5cm}\forall k \in \intset{0,T}, \forall i \in \intset{1,q}, \forall v \in P_H, \forall p \in \intset{1,n} \setminus v \\
    d^k_{i,v} \geqslant \underset{p \in v}{\sum} b^k_{i,p} + \underset{\substack{p=1 \\ p \notin v}}{\overset{n}{\sum}} (1-b^k_{i,p}) + 1 - n& \\  &\hspace*{-3cm}\forall k \in \intset{0,T}, \forall i \intset{1,q}, \forall v \in P_H.
    \end{cases}
\end{equation}

\textbf{Rotation validity:} a rotation corresponding to a given monomial can only occur if that monomial is properly represented on a given qubit,
\begin{equation}
    r^k_{i,v} \leqslant d^{k-1}_{i,v} \quad \forall k \in \intset{1,T}, \forall i \in \intset{1,q}, \forall v \in P_H.
\end{equation}

\textbf{Propagation rule:} modelling the behavior of CNOT gates, and the overall propagation of terms throughout the circuit, can be best explained as a binary sum $\oplus_2$, 
\begin{multline*}
    b^k_{i,p} = b^{k-1}_{i,p} \oplus_2 (\underset{j=1}{\overset{q}{\sum}} c^k_{j,i} b^{k-1}_{j,p}) \\ \forall k \in \intset{1,T}, \forall i \in \intset{1,q}, \forall p \in \intset{1,n}.
\end{multline*}

Note that $\underset{j=1}{\overset{q}{\sum}} c^k_{j,i} b^{k-1}_{j,p}$ is indeed a binary quantity, due to the $b$ variables being binary and to the uniqueness constraint. This binary sum can be rewritten in the following manner:
\begin{equation}
    b^k_{i,p} = b^{k-1}_{i,p} + \underset{j=1}{\overset{q}{\sum}} c^k_{j,i} b^{k-1}_{j,p} - 2 b^{k-1}_{i,p} \underset{j=1}{\overset{q}{\sum}} c^k_{j,i} b^{k-1}_{j,p}.
    \label{eq:prop_rule}
\end{equation}

This expression can then be linearized at the cost of introducing additional variables.

\subsection{The routing problem}

The circuit design problem (CDP) can then be formulated as a special routing problem: we route the singletons $Z_i$ through the circuit, using their initial positions and CNOT gates, in order to compute each monomial of the Hamiltonian, once in the circuit. As said, the goal is to minimize the depth of the circuit, subject to all previously described constraints, modelling the initial positions of singletons, their routing through CNOT gates, and the presence of a phase rotation gate for each monomial of the Hamiltonian. Tthe problem reads, 
\begin{equation}\label{full:opt} (\mathrm{CDP})\hspace{1mm}
    \begin{cases}
    \underset{\footnotesize\begin{array}{c} a \in \{0,1\}^T, b\in \{0,1\}^{nqT},\\ c\in \{0,1\}^{q^2T}, d, r \in \{0,1\}^{qT |P_H|}
    \end{array}}{\min} J(\{a^k\}) \\
    \hspace*{2cm} \text{subject to} \hspace{3.5mm} (\ref{eq:init}) - (\ref{eq:prop_rule}).
    \end{cases}
\end{equation}

We immediately note that variables $a$, $b$ and $d$ can be set to continuous in $[0,1]$ instead of binary without loss of generality. In fact, if $c$ and $r$ take binary values, then the constraints imply that all other variables also take binary values.

The constraints can also be augmented with trivial symmetry-breaking conditions, such as $a^k \geqslant a^{k+1}$. This stacks any activity in the circuit on the preceding layers, thus not leaving any gaps in the circuit.

As one can expect, the (CDP) routing problem is difficult to solve to optimality in reasonable amounts of time. Even for small Ising Hamiltonians with relatively few variables and monomials, one will have to resort to solver heuristics. For example, working with Gurobi as a solver, we are able to solve it in seconds when $f$ is a single monomial of degree three. When we raise it to degree four, optimality is out of reach, even allowing for several hours of computational time. Then, it is key to devise simplifications and heuristics to tackle the problem fast yet efficiently. 

\section{Simplifications and heuristics}\label{sec:solutions}

As explained in the previous section, solving (CDP) to optimality is often impossible given a reasonable time limit. However, the formulation is still useful in a variety of ways. First, by using solver heuristics, we can generate circuits that can improve on feasible solutions given as a warm start (e.g., a Qiskit-generated compilation). Second, imposing further constraints to reduce the search space, we can trade off optimality for faster resolution.  

\subsection{Simplifications: downward CNOTs}

One such set of constraints is remarkably simple yet very effective, and it can also be motivated by current quantum technology. If we constrain all the CNOT gates to be all in the same direction, e.g., downward facing, then the search space is reduced enough to allow us to compute optimal solutions more efficiently. The constraint reads,
\begin{equation}\label{eq:cnot-down}
    c^k_{i,j} = 0 \quad \forall k \in \intset{1,T},\hspace{1mm} \forall i > j \in \intset{1,q}^2.
\end{equation}
This constraint imposes that singletons $Z_i$ can only travel ``downwards", to qubits with a higher index. This constraint is somewhat natural, it is often a constraint in current quantum technology, and it appears in standard compilation tools (e.g., Qiskit orients its CNOTs), see also~\cite{madden2022best}. Again, working with Gurobi, this enables us to solve (CDP) when $f$ is a monomial of degree four in seconds, and even if the solution obtained is not optimal in terms of depth (as we will show in the result section), the result is better than Qiskit compilation. 

\subsection{Templates}

\begin{figure*}
\centering
    \centering
    \includegraphics[width=0.725\textwidth]{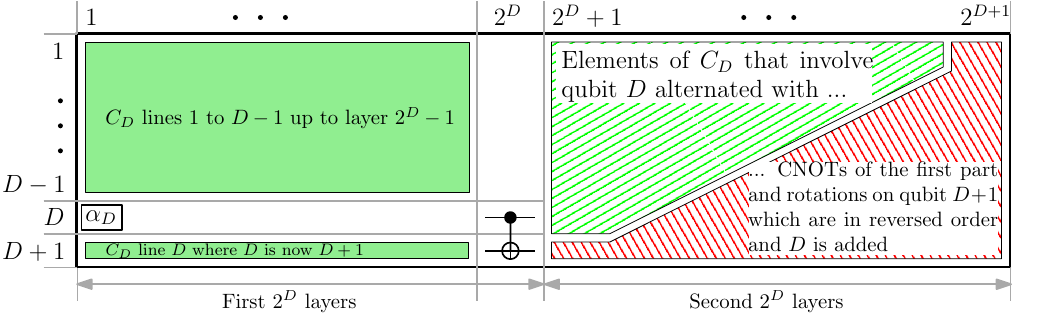}
    \caption{Visual explanation of Algortihm~\ref{alg:cap}. Consider $C_D$ as a rectangle $\mathcal R =((1,1)\to(D, 2^D))$. We build $C_{D+1}$ as a rectangle $\mathcal R'=((1,1)\to(D+1, 2^{D+1}))$ with:
{\bf first column} of $\mathcal R'$ has the rotations on the $D+1$ singletons; 
{\bf column} $2^D$ of $\mathcal R'$ has a CNOT from $D$ to $D+1$;
{\bf rectangle} $\mathcal R''=((1,1)\to(D-1, 2^D -1))$ contains $C_D$ lines $1$ to $D-1$ up to layer $2^D-1$; 
{\bf rectangle} $\mathcal R'''=((1,2^{D}+1)\to(D+1, 2^{D+1}))$ contains two intertwined circuits: first the elements of $C_D$ that involve qubit $D$ alternated with the CNOTs of the first part and rotations on qubit $D+1$ which are in reversed order and $D$ is added. 
}
    \label{fig:explanation}
\end{figure*}

We now explore this simplification further, to devise a recursive template (i.e., a circuit that can be used for any monomial of a given degree) that can be generated by an algorithm in polynomial time, which is able to synthetize the circuit of any HOBO problem of $m$ monomials and degree $D$ in $O(m \, 2^D)$ layers. This is key to the practical implementability of (CDP), and the resulting circuit is still of reasonable quality.  This template was found by computer-assisted trial-and-error, by using the simplified problem (CDP) with~\eqref{eq:cnot-down}, by fixing $T = 2^D$ and minimizing the number of CNOT gates. The resulting template is far from being trivial, so its generation may seem convoluted at first. Yet, it is surprisingly very effective. 

Let us consider the design of a template for a circuit without ancillas for $f$ when $f$ is a monomial of a given degree $D$. This circuit requires $D$ qubits.  
Let us denote this circuit by $C_D$. We can start with $C_2$ (i.e., a QUBO), which we can solve at optimality and it is given in Figure \ref{fig:template}a: a circuit of depth equal to $4$.  
Imagine now that we have access to $C_D$, $D\geq2$, we can now generate $C_{D+1}$ as follows. 

\begin{algorithm}
\caption{Template generator}\label{alg:cap}
\small
\begin{algorithmic}[1]
\Require  A template for $C_D$ in $D$ qubit and $2^D$ layers. 
\Ensure A template for $C_{D+1}$ in $D+1$ qubit and $2^{D+1}$ layers.
\State Add all the phase rotation gates $\alpha_q$, $q \in \intset{1,D+1}$ on the first layer
\State Add layers from $2$ to $2^{D-1}$ of $C_D$ applied to qubits $q\in\intset{1,D-1}\cup\{D+1\}$
\State For layer $2^D$, add a CNOT gate controlled by qubit $D$ acting on qubit $D+1$
\State For layers $2^D+1$ till $2^{D+1}$ alternate \emph{(i)} $C_D$ applied to qubits $\intset{1,D}$ with rotations only on qubit $D$, and \emph{(ii)} the CNOT gates of the first $2^D$ layers with rotations on qubit $D+1$. These latter rotations are in number $2^D$ with element $D$ added and otherwise taken in reversed order with respect to the first $2^D$ layers.
\end{algorithmic}
\end{algorithm}

We further expand on the explanation of the template generator in Figure~\ref{fig:explanation}, and its caption, as well as in Figure~\ref{fig:template}b and Figure~\ref{fig:template}c. Note that in the latter figures, we superpose rotation gates to CNOT gates (in somewhat a non-traditional layout) to stress that the rotation and the CNOT happen in parallel on the same layer. 

We are now ready for the following proposition. 

\smallskip
\begin{proposition}\label{prop:1}
The Template generator in Algorithm~\ref{alg:cap} outputs a circuit $C_{D+1}$ given the preceding circuit $C_D$ that has all the rotation gates we need to represent a monomial of degree $D+1$, it verifies all the constraints of (CDP) and it has exactly $2^{D+1}$ layers and $2^{D+1}$ unique rotation gates.
\end{proposition}
\smallskip

\small
{\bf Proof.} To represent a monomial of degree $D+1$, we need all the unique combination of indices up to degree $D+1$, of which there are $2^{D+1}-1$. With our construction, $C_{D+1}$ has $2^{D}-1$ rotations coming from $C_D$ plus one considering $\alpha_D$. To this we add all the rotations on the second $2^{D}$ layers, i.e., all the rotations on qubit $D$, as $2^{D-1}-1$ and all the ones on qubit $D+1$, as $2^{D-1}$. Adding up we obtain a total of unique rotation gates of $2^{D+1}-1$, as required. 

As for the layers: $C_{D+1}$ has firstly $2^{D}$ layers by construction, at which we add all the CNOTs plus rotations of these first $2^{D}$ layers a second time, adding another $2^D$ layers, totalling at $2^{D+1}$. Note that the circuit acting on qubit $D$ on the second half can happen in parallel with the one acting on qubit $D+1$, so it is not counted here. 


Finally, one can verify that all the constraints of (CDP) are imposed by construction and they hold. 
\hfill  \qed

\normalsize

\smallskip
Proposition~\ref{prop:1} suggests that our template generator is also \emph{theoretically} quite efficient, using the monomial structure to compile its resulting unitary matrix on $D$ qubits in $2^{D}$ layers. This is in contrast with the $O(4^D)$ lower bound for generic unitary matrices~\cite{shende2004lowerbound}. When considering a degree $D$ monomial in binary variables, since the reformulation creates $2^D-1$ terms to encode, the scaling in $O(2^D)$ may be unavoidable.

\begin{figure*}
\centering
    \begin{subfigure}{0.28\textwidth}
    \centering
    \includegraphics[width=1\textwidth]{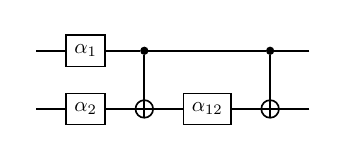}
    \subcaption{Template circuit $C_2$.}
    \end{subfigure}
    \hfill 
    \begin{subfigure}{0.51\textwidth}
    \includegraphics[width=1\textwidth]{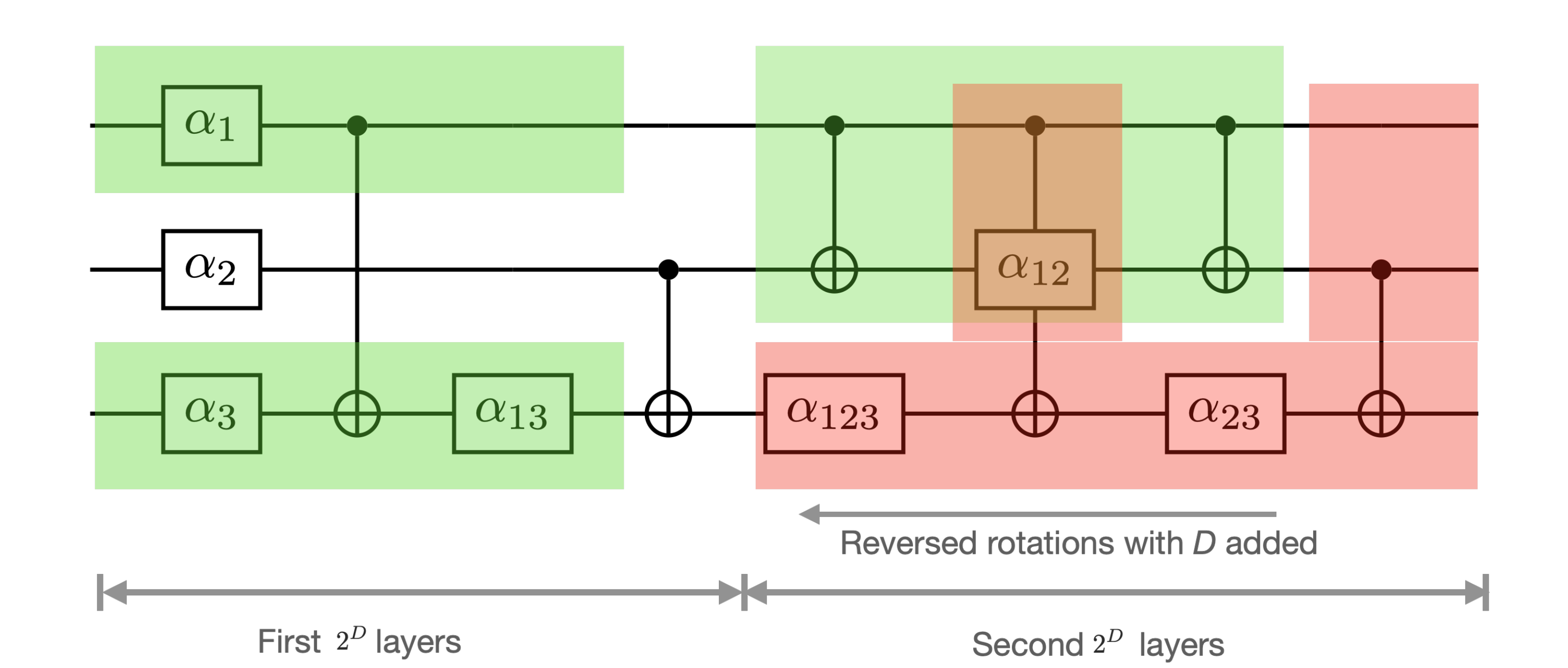}
    \subcaption{Template circuit $C_3$, with the color code from the algorithm explanation.}
    \end{subfigure}
    \vskip 10pt
    \begin{subfigure}{0.88\textwidth}
    \includegraphics[width=1\textwidth]{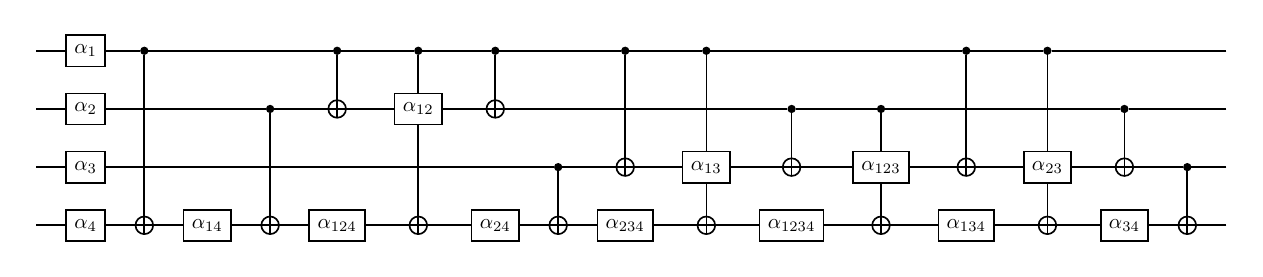}
    \subcaption{Template circuit $C_4$.}
    \end{subfigure}
    \caption{Template circuits for binary monomials of $D=2, 3, 4$. Note the non-standard visual layout of the superposition of rotation and CNOT gate (e.g., $\alpha_{12}$ in $C_3$), which we use to stress that both gates are applied in the same layer. }
    \label{fig:template}
\end{figure*}

\begin{figure*}
\setcounter{figure}{0}    
\renewcommand{\figurename}{Table}
\centering
\caption{Sample results for the considered instances, showcasing the benefit of our methodologies. We have imposed a $1200$~s time limit. In the case of $-$, the solver could not improve on the warm start solution or prove its optimality in the time limit.}
\label{tab:results}
\centering
\scalebox{0.76}{
\begin{tabular}{cccccccccccccccccc}
	\toprule
    & &&&\multicolumn{2}{c}{Qiskit \cite{Qiskit2019}} & & \multicolumn{2}{c}{Gray code \cite{HOBO}} & & \multicolumn{2}{c}{(CDP) \eqref{full:opt}}& & \multicolumn{2}{c}{Simpl.ed (CDP) \eqref{full:opt} + \eqref{eq:cnot-down}}& & \multicolumn{2}{c}{Templates, Algo.~\ref{alg:cap}}\\ \cmidrule(lr){5-6}\cmidrule(lr){8-9}\cmidrule(lr){11-12}\cmidrule(lr){14-15}\cmidrule(lr){17-18}
	Instance & $n$ & $|P|$ & $D$ & Depth & CPU [s] & & Depth & CPU [s] & & Depth & CPU [s] & & Depth & CPU [s] & & Depth & CPU [s] \\
	\toprule
    qubo1  & $4$ & $5$ & $2$ & 17 & 0.6 &    & 35 & $<$0.1 &    & {\bf 8} & 481 &    & 8 & 0.3 &    & 12 & $<$0.1\\
    qubo2 & $6$ & $10$ & $2$ & 29 & 0.6 &    & 70 & $<$0.1 &    & 16 & 1200 &    & {\bf 11} & 6.9 &    & 20 & $<$0.1\\
    \midrule
    monomial3 & $3$ & $1$ & $3$ & 12 & 0.5 &    & 15 & $<$0.1 &    & {\bf 8} & 14 &    & {\bf 8} & 0.1 &    & {\bf 8} & $<$0.1\\
    poly3-1 & $4$ & $2$ & $3$ & 20 & 0.6 &    & 30 & $<$0.1 &    & {\bf 9} & 1200 &    & 10 & 1.6 &    & 16 & $<$0.1\\
    poly3-2 & $5$ & $4$ & $3$ & 41 & 0.6 &    & 60 & $<$0.1 &    & 26 & 1200 &    & {\bf 16} & 8.0 &    & 32 & $<$0.1\\
    \midrule
    monomial4 & $4$ & $1$ & $4$ & 26 & 0.6 &    & 31 & $<$0.1 &    & {\bf 16} & 1200 &    & {\bf 16}& 2.4 &    & 16 & $<$0.1\\
    poly4-1 & $5$ & $2$ & $4$ & 45 & 0.6 &    & 62 & $<$0.1 &    & 26 & 1200 &    & {\bf 18} & 1200 &    & 32 & $<$0.1\\
    poly4-2 & $6$ & $4$ & $4$ & 89 & 0.8 &    & 124 & $<$0.1 &    & - & - &    & {\bf 47} & 1200 &    & 64 & $<$0.1\\
    \midrule
    monomial5 & $5$ & $1$ & $5$ & 56 & 0.6 &    & 63 & $<$0.1 &    & - & - &    & - & - &   & {\bf 32} & $<$0.1\\
    poly5-1 & $6$ & $2$ & $5$ & 101 & 0.7 &    & 126 & $<$0.1 &    & - & - &    & {\bf 57} & 1200 &   & 64 & $<$0.1\\
    poly5-2 & $7$ & $4$ & $5$ & 206 & 1.1 &    & 252 & $<$0.1 &    & - & - &    & - & - &   & {\bf 128} & $<$0.1\\
    \midrule
    monomial6 & $6$ & $1$ & $6$ & 118 & 1.7 &    & 127 & $<$0.1 &    & - & - &    & - & - &   & {\bf 64} & $<$0.1\\
    poly6-1 & $7$ & $2$ & $6$ & 228 & 1.3 &    & 254 & $<$0.1 &    & - & - &    & - & - &   & {\bf 128} & $<$0.1\\
    poly6-2 & $8$ & $4$ & $6$ & 438 & 1.9 &   & 508 & $<$0.1 &    & - & - &    & - & - &   & {\bf 256} & $<$0.1\\
	\bottomrule
\end{tabular}
}
\end{figure*}

We can now use the template generator to devise a simple heuristic to design a variational circuit for a given HOBO problem. The idea is to consider all untreated monomials, find out which monomial has its corresponding qubits available at the earliest point in the circuit, and add to the circuit the template corresponding to the chosen monomial. Repeat this procedure until all monomials have been treated. This heuristic is simple, compiling a circuit featuring $m$ monomials up to degree $D$ in $O(m\,2^D)$ layers, and effective in practice. Of course the $O(m\,2^D)$ is only an upper bound, especially when monomials pertain a disjoint set of qubits and they can be treated simultaneously, as well as other possible simplifications, which will be considered in future work.

\section{Results}\label{sec:results}

We compare the depth of circuits obtained on few instances with different methods. We use the combinatorial optimization (CDP)~\eqref{full:opt} described in Section \ref{sec:formulation}, its simplified version adding the constraints~\eqref{eq:cnot-down}, the template generator heuristic of Algorithm~\ref{alg:cap}, Qiskit~\cite{Qiskit2019}, and the Gray code approach of~\cite{HOBO}. In the case of Qiskit, we design the circuit as if multiple-qubit gates existed, and then use a pass manager to unroll the circuit into one-qubit and CNOT gates only, and then use a second pass to optimize the design.

We consider various monomial and polynomial instances of varying degree, from QUBOs of degree two, to polynomials of degree six. In all cases, we report the obtained depth and used computational resources on a laptop machine with Intel Core i7-10810U processors (4.9 GHz) with 16 GB of RAM. For (CDP) and its simplified version, we use Gurobi~\cite{gurobi} as a mixed-integer linear programming solver, we give the circuit obtained with the template generator as feasible starting solution, and we impose a $1200$~s time limit. If the time limit is reached, optimality is therefore not guaranteed.

Table~\ref{tab:results} shows that our approaches outperform Qiskit and the Gray code method of \cite{HOBO} (which even uses an ancilla qubit) in all the instances. Given the time limit, the simplified strategy can be better than (CDP), and our template is surprisingly very effective in delivering more compact circuits in less than $0.1$ seconds. Future research will explore our template more thoroughly, and mix it with other ideas for heuristics.

\section{Conclusion}\label{sec:conclusion}

We have studied how to compile variational circuits stemming from higher-order polynomial binary optimization (HOBO) problems. We have formulated the compilation task as a qubit routing combinatorial problem and proposed ways to solve it efficiently. In particular, we have found templates that can compile any HOBO with a recursive algorithm in polynomial time, sacrificing minimal depth circuits for almost immediate compilation. The results show that our approach is effective and promising. Remaining questions include the resulting circuit's resilience to noise, which may be studied in future work.

\section*{Acknowledgements}
This research benefited from the support of the FMJH Program PGMO, under project number 2022-0010. 

\bibliographystyle{IEEEtran}
\bibliography{IEEEabrv,bibliography}
\end{document}